\documentclass[letter]{elsarticle}

\usepackage{graphicx,amsmath,amssymb,bm,pdfpages,caption,lineno,hyperref}
\modulolinenumbers[5]

\journal{Physics Letters B}

\newcommand{\be}{\begin{equation}}
\newcommand{\ee}{\end{equation}}

\newcommand{\mev}{\, \text{MeV}}
\newcommand{\tev}{\, \text{TeV}}
\newcommand{\kev}{\, \text{keV}}

\begin{document}
\begin{frontmatter}
\title{Beta spectrum of unique first-forbidden decays as a novel test for fundamental symmetries}

\author{Ayala\ Glick-Magid}
\address{Racah Institute of Physics, 
The Hebrew University, 
9190401 Jerusalem, Israel}
\author{Yonatan\ Mishnayot}
\address{Racah Institute of Physics, 
The Hebrew University, 
9190401 Jerusalem, Israel}
\address{Department of Particle Physics and Astrophysics, The Weizmann Institute, Rehovot 7610001, Israel}
\address{Soreq Nuclear Research Center, Yavne 81800, Israel}
\author{Ish\ Mukul}
\address{Department of Particle Physics and Astrophysics, The Weizmann Institute, Rehovot 7610001, Israel}
\author{Michael\ Hass}
\address{Department of Particle Physics and Astrophysics, The Weizmann Institute, Rehovot 7610001, Israel}
\author{Sergey\ Vaintraub}
\address{Soreq Nuclear Research Center, Yavne 81800, Israel}
\author{Guy\ Ron}
\address{Racah Institute of Physics, 
The Hebrew University, 
9190401 Jerusalem, Israel}

\author{Doron\ Gazit}
\cortext[mycorrespondingauthor]{Corresponding author}
\ead{doron.gazit@mail.huji.ac.il}
\address{Racah Institute of Physics, 
The Hebrew University, 
9190401 Jerusalem, Israel}

\begin{abstract}
Within the Standard Model, the weak interaction of quarks and leptons is characterized by certain symmetry properties, such as maximal breaking of parity and favored helicity. These are related to the $V-A$ structure of the weak interaction.
These characteristics were discovered by studying correlations in the directions of the outgoing leptons in nuclear beta decays. Presently, correlation measurements in nuclear beta decays are intensively studied to probe for signatures for deviations from these couplings, which are an indication of Beyond Standard Model physics. 
We show that the structure of the energy spectrum of emitted electrons in unique first-forbidden $\beta$-decays is sensitive to the symmetries of the weak interaction, and thus can be used as a novel probe of physics beyond the standard model. Furthermore, the energy spectrum gives constraints both in the case of right and left couplings of the new beyond standard model currents. 
We show that a measurement with modest energy resolution of $\approx 20 \kev$ is expected to lead to new constraints on beyond the standard model interactions with tensor couplings.
\end{abstract}


\end{frontmatter}

Since their discovery, nuclear $\beta$ decays have been used as heralds of new physics. The existence of the neutrino was conjectured by Pauli, using the continuous spectrum of the electron. Later, experiments emphasized the study  of the kinematics of $\beta$ decays, in particular the angular correlation between the directions of the emitted $\beta$ particle, i.e., electron or positron, and the neutrino ($\nu$) (or, equivalently, the recoiling nucleus)~\cite{PhysRev.106.517}. In particular, the famous Wu experiment~\cite{PhysRev.105.1413} has proven the breaking of parity symmetry. In recent years, several experiments  are using precision measurements of  the $\beta$-$\nu$ 
angular correlation coefficient in the decay of the short-lived radio-nuclides (see e.g.~\cite{AnnRevNuclPartPhys.61.23,RevModPhys.78.991,AnnalenPhys.525.600,PhysScripta.T152.014018,RevModPhys.87.1483} and references therein). These experiments search for the minute experimental signal that originates from possible tensor or scalar terms in the weak interaction. Such scalar or tensor terms modify the angular correlation between a neutrino and an electron in the beta-decay process, thus probing new physics of Òbeyond-the-standard-modelÓ (BSM) nature~\cite{JPhysG.41.110301,ProgPartNuclPhys.71.93}. Present limits on possible deviations from the standard model predictions are of the order of $0.1-1$\% \cite{AnnalenPhys.525.600,RevModPhys.87.1483,PhysRevLett.115.162001}, broadly yielding a limit on the scale of new physics of the order of $\sim 1-10 \tev$ \cite{NuclPhysB.830.95,JHEP.2.1}. The latter limits originate in analysis of $\beta$-decays, assuming BSM couplings to left handed neutrinos, and become significantly worse for right handed neutrinos.
 
Precision measurements which are based on correlations of the emitted leptons have several disadvantages. In particular, most modern experiments make use 
of trapped ions or atoms in order to characterize the kinematics. The use of traps allows a significant reduction in the systematic uncertainties enabling precise correlation
experiments. This limits the number of possible radio isotopes, as trapping is efficient for relatively short times. Additionally, complex detection and analysis schemes are required for the extraction of the correlations~\cite{JPhysG.41.110301,JPhysG.41.114005}. Current experiments mostly study allowed $\beta$-decays, where constraining separately right handed and left handed couplings is currently out of reach, due to inconsistencies in the energy averaging (See Ref.~\cite{PhysRevC.94.035503} for details).

Here, we propose a novel probe for beyond the standard model couplings, the energy spectrum 
of $\beta$ decays that are characterized by transitions in which the total angular momentum in the daughter and mother nuclei differ by $\Delta J=2$ units, as well as by change in parity. These decays are commonly known as {\it unique first-forbidden decays}. Such measurements are potentially simpler than a precise determination of the $\beta$-$\nu$ correlation coefficient, demand neither trapping nor cooling, and require a single observable to characterize. Additionally, a measurement of the full $\beta$ spectrum, is less amenable to detector calibration and resolution issues, which plague $beta$ endpoint measurements. More importantly, they provide constraints on exotic couplings and thus allow identifying systematic errors, a valuable feature for precision studies.  

In order to show the difference between allowed and unique first-forbidden decays, it is convenient to write the general differential distribution of $\beta$-electron (positron) of energy $\epsilon$, momentum $\vec{k}$ and direction $\vec{\beta}=\frac{\vec{k}}{\epsilon}$,  and neutrino $\bar{\nu}(\nu)$ of momentum $\vec{\nu}$ in a $\beta^{\mp}$ decay process, as follows:
\begin{equation} \label{Eq:Separation}
\frac{d^5\omega_{\beta^{\mp}}}{d\Omega_k/4\pi d\Omega_\nu/4\pi d\epsilon} = \Sigma(\epsilon) \cdot \Theta(q, \vec{\beta}\cdot\hat{\nu}).
\end{equation}
With $\vec{q}=\vec{k}+\vec{\nu}$ is the momentum transfer in the process.

$\Sigma(\epsilon)$ is a nuclear independent part, related to the electrostatic interaction between the $\beta$ particle and the decaying nucleus,
\begin{equation} \label{Eq:Sigma_eps}
\Sigma(\epsilon)=\frac{2G^2}{\pi^2} 
\frac{2\Delta J +1}{\Delta J(2J_i+1)}(\epsilon_{0}-\epsilon)^2 k\epsilon\, F^{(\pm)}(Z_f,\epsilon),
\end{equation}
with $G$ the Fermi constant, $J_i$ is the total angular momentum of the decaying (mother) nucleus, $\Delta J$ is the difference between the angular momenta of the mother and daughter nuclei, $Z_f$ is the charge of the daughter nucleus,  and $\epsilon_{0}=\frac{2Q+Q^2+m_{e}^2}{2Q+2m_{r}}$ \cite{RevModPhys.46.789} is the maximum electron energy ($m_e$ and $m_r$ are electron and daughter nucleus masses, respectively, $Q$ is the decay $Q$-value). 
The deformation of the lepton wave
function due to the long--range electromagnetic interaction with the
nucleus is taken into account in the Fermi
function $F^{(\pm)}$ for a $\beta^{(\pm)}$ decay \cite{ZPhysik.88.161,NuovoCim.11.1},
\begin{equation}
  F^{(\pm)}(Z_f,\epsilon)=2(1+\gamma_0) (2 \epsilon R_f)^{2(\gamma_0-1)}
  \frac{\mid \Gamma(\gamma_0+i\rho) \mid^2}{\mid \Gamma(2\gamma_0+1)
  \mid^2} e^{\pi \rho}
\end{equation}
with $\alpha \approx 1/137$ the fine structure constant, $R_f$
the radius of the final nucleus, $\rho= \mp \alpha Z_f / \beta_f$ ($\beta_f$ is the momentum to energy ratio of the $\beta$ particle),
and $\gamma_0=\sqrt{1-(\alpha Z_f)^2}$ ($\Gamma(x)$ is the Gamma function).

Assuming the Standard Model ($V-A$) coupling, the second term in Eq.~(\ref{Eq:Separation}), i.e., the function $\Theta(q,\vec{\beta}\cdot\hat{\nu})$, depends on the  nuclear wave functions, and is usually written using a multipole expansion \cite{WaleckaBook1995},

\begin{eqnarray} \label{Eq:general_beta_decay}
\Theta(q,\vec{\beta}\cdot\hat{\nu}) &=&{ \frac{\Delta J}{2\Delta J+1} \left\{ \left[ 1-\bigl(\hat{\nu} \cdot \hat{q}
\bigr) \bigl(\vec{\beta} \cdot \hat{q} \bigr) \right] \sum_{J \ge
1} \bigl(|\langle\|{\hat{E}_J} \|\rangle|^2+ |\langle \|{\hat{M}_J} \|\rangle|^2 \bigr)\right.\pm}\\ \nonumber
 & &\pm{\left. \hat{q} \cdot \left( \hat{\nu} - \vec{\beta} \right) \sum_{J
\ge 1} 2 \Re \langle \|{\hat{E}_J} \|\rangle\langle \|{\hat{M}_J} \|\rangle^{*} + \right. }\\
\nonumber & &{+ \left. \sum_{J \ge 0} \left[ \left[ 1-\hat{\nu}
\cdot \vec {\beta}+2\bigl(\hat{\nu} \cdot \hat{q} \bigr)
\bigl(\vec{\beta} \cdot \hat{q} \bigr) \right] |\langle \|{\hat{L}_J} \|\rangle|^2 + \left( 1+\hat{\nu} \cdot \vec {\beta} \right)
|\langle \|{\hat{C}_J} \|\rangle|^2 - \right.\right.}\\ \nonumber
&&{-\left.\left. 2 \hat{q} \cdot \left( \hat{\nu} +
\vec{\beta} \right) \Re \langle \|{\hat{C}_J} \|\rangle\langle \|{\hat{L}_J} \|\rangle^{*}
\right] \right\} },
\end{eqnarray}
where, $\langle \|{\hat{O}}_J \|\rangle$, is the reduced matrix element of a rank $J$ spherical tensor operator $\hat{O}_J$, between the daughter and mother wave functions. 

The multipole operator decomposition of the nuclear current, viz. the Coulomb, electric, magnetic, and longitudinal
operators:
\begin{align}
\hat{{C}}_{JM}(q) & =  \int {d\vec{x}
j_{J}(qx)Y_{JM}(\hat{x})\hat{\mathcal{J}}_{0}(\vec{x})}
\\
\hat{{E}}_{JM}(q) & =\frac{1}{q}\int{d\vec{x}\vec {\nabla} \times
[j_J(qx)\vec{Y}_{JJM}(\hat{x})]\cdot
\hat{\vec{\mathcal{J}}}(\vec{x})}
\\
\hat{{M}}_{JM}(q) & =\int{d\vec{x}j_J(qx)\vec{Y}_{JJM}(\hat{x})\cdot
\hat{\vec{{\mathcal{J}}}}(\vec{x})}
\\
\hat{L}_{JM}(q) & =\frac{i}{q}\int{d\vec{x}\vec {\nabla}
[j_J(qx){Y}_{JM}(\hat{x})]\cdot \hat{\vec{\mathcal{J}}} (\vec{x})},
\end{align}
where $\hat{{\mathcal{J}}^\mu} (\vec{x})$ is the nuclear current coupling to the probe. For $\beta$-decays, which are characterized by a low-energy transfer $qR \ll 1$, further simplification is possible expanding in this small parameter. For example, for allowed $\beta$-decays with $\Delta J^\pi=1^{+}$ (Gamow-Teller decays), 
\begin{equation}
\Theta \propto (1+b\frac{m_e}{\epsilon}+a_{\beta\nu}\vec{\beta}\cdot\hat{\nu})\langle \|{\hat{L}}_1\|\rangle^2,
\end{equation}
where $m_e$ is the electron mass. This is accurate up to (recoil) corrections of order $qR$. The $V-A$ structure of the weak interaction entails ${a}_{\beta\nu}=-\frac{1}{3}$ and $b=0$. In the presence of beyond standard model interaction with tensor symmetry ${a}_{\beta\nu}\approx-\frac{1}{3}\bigl({1-\frac{|C_T|^2+|C_T'|^2}{|C_A|^2}}\bigr)$, and $b=2\frac{C_T+C_T'}{C_A}$~\cite{PhysRev.106.517}, where $C_T/C_A$ ($C_T'/C_A$) is the relative strength of the tensor (pseudo-tensor) and the axial-vector interactions \footnote{For simplicity, we assume here real couplings, i.e., time reversal symmetry}. Thus, $\beta$-$\nu$ correlation measurements are sensitive to interactions of exotic, e.g., tensor, symmetries. However, the $\beta$ energy spectrum form shows sensitivity only to the Fierz interference term $b$, since
\begin{equation}
\frac{d\omega_{\beta^\mp}}{d\epsilon}(allowed)\propto \Sigma(\epsilon) \bigl(1+b\frac{m_e}{\epsilon}\bigr).
\end{equation}
The Fierz term is linear in the exotic couplings, while $a_{\beta\nu}$ is quadratic. In addition, the Fierz term vanishes for right-handed neutrinos, for which $C_T=-C_T'$. As a result, allowed $\beta$ decay measurements are better able to constrain the combination $C_T+C_T'$. Moreover, current experiments cannot fit separately both $a_{\beta\nu}$ and the Fierz term 
(even when including a non-zero Fierz term in the analysis) \cite{PhysRevC.94.035503},
 due to the fact that the correlation and Fierz terms have different recoil momentum dependence.
This,  however, is not the case for a first forbidden unique transition, where a different result is obtained (see, e.g., \cite{PhysRev.82.531}),
\begin{eqnarray} \label{Eq:Angle_part_first_forbidden}
\Theta(q,\vec{\beta} \cdot \hat{\nu})&\propto& 1 \pm2\gamma_0\frac{C_T+C_T'}{C_A}\frac{m_e}{\epsilon}- \\ \nonumber 
&-&\frac{1}{5}\bigl(2\bigl(\hat{\nu}
\cdot \vec {\beta}\bigr)-\bigl(\hat{\nu} \cdot \hat{q} \bigr)\bigl(\vec{\beta}\cdot\hat{q}\bigr)\bigr)
\bigl({1-\frac{|C_T|^2+|C_T'|^2}{|C_A|^2}}\bigr) .
\end{eqnarray}
The last term linearly depends on $\bigl(\hat{\nu}
\cdot \hat {k}\bigr)^2$.
As a result, integration over angles, i.e., the energy spectrum of the decay, is sensitive to beyond the standard model tensor interactions, and can be used to probe them. Moreover, a full spectrum measurement enables a simultaneous extraction of $C_T+C_T'$ and $C_T-C_T'$ is possible, allowing studies of right and left handed neutrino couplings. 

{\it Theory related systematic corrections--} In the derivation of Eq.~\ref{Eq:Angle_part_first_forbidden}, we have used an expansion in $qR$. The neglected recoil corrections terms compete with signatures of tensor and/or other beyond the standard model contributions. In order to estimate the recoil corrections, let us write the decay rate of a unique first-forbidden decay up to the next-to-leading order in the parameter $qR$ (assuming no tensor terms):
\begin{eqnarray} \label{Eq:first_forbidden_beta_decay}
\frac{d^5\omega_{\beta^{\mp}}}{d\Omega_k/4\pi d\Omega_\nu/4\pi d\epsilon} &=& \frac{2G^2}{\pi^2} 
\frac{1}{2J_i+1}(\epsilon_{0}-\epsilon)^2 k\epsilon\, F^{\pm}(Z_f,\epsilon)
 \times \nonumber \\  \nonumber 
&&{ \times \left\{ \frac{5}{2}\left[ 1+\delta_1 -\frac{2}{5}(1+\delta_{\hat{\nu}\cdot\vec{\beta}})\hat{\nu}
\cdot \vec {\beta} \right.\right.} +\\  &&+{\left.\left.\frac{1}{5}\bigl(\hat{\nu} \cdot \hat{q} \bigr)
\bigl(\vec{\beta} \cdot \hat{q} \bigr) \right] \langle \|{\hat{L}^A_2} \|\rangle^2  \right\}},
\end{eqnarray} 
with,
\begin{align}
\delta_1 &= \frac{4}{5} \left\{\pm\sqrt{\frac{3}{2}}\frac{\nu-\frac{k^2}{\epsilon}}{q} \Re \frac{\langle \|{\hat{M}^V_2} \|\rangle}{\langle \|{\hat{L}^A_2} \|\rangle}-\frac{\nu+\frac{k^2}{\epsilon}}{q} \Re \frac{\langle \|{\hat{C}^A_2} \|\rangle}{\langle \|{\hat{L}^A_2} \|\rangle} \right\}
\\
\delta_{\hat{\nu}\cdot\vec{\beta}} &=2 \left\{\pm\sqrt{\frac{3}{2}}\frac{\epsilon-\nu}{q} \Re \frac{\langle \|{\hat{M}^V_2} \|\rangle}{\langle \|{\hat{L}^A_2} \|\rangle}-\frac{\nu+{\epsilon}}{q} \Re\frac{\langle \|{\hat{C}^A_2} \|\rangle}{\langle \|{\hat{L}^A_2} \|\rangle} \right\}
\end{align}
where the superscript $A$ ($V$) denotes multipole operators calculated with the axial-vector (polar-vector) symmetry contribution to the weak nuclear current. Ordering the multipoles by their $qR$ dependence, we see that $\hat{L}^A_2$ is ${\mathcal{O}}(qR)$, while $\hat{C}_2^A,\,\hat{M}_2^V$ are suppressed by an additional factor of $qR$, which for relevant $Q$-values of a unique first-forbidden decay, i.e., $Q\approx 10\mev$, leads to a factor of 20. Moreover, as $j_J(\rho)\sim  \frac{\rho^J}{(2J+1)!!}$ (for $\rho\ll 1$), we find an additional 
suppression factor of 5.
An important aspect for estimating the neglected recoil corrections, originates in the fact that the nuclear weak current can be organized perturbatively using chiral effective field theory. We keep only leading and next-to leading order. To this order, the weak probe interacts with a single nucleon, such that:
${\mathcal J}^{\mu\dagger}({\bf r}) = \sum_{i=1}^{A} \tau_{i}^{-} \bigl[
\delta^{\mu0} J_{i,{\rm 1b}}^{0} - \delta^{\mu k}J_{i,{\rm 1b}}^{k}
\bigr] \delta({\bf r}-{\bf r}_i)$,
where $\tau^{-}=\frac{1}{2}(\tau^{x}-i\tau^{y})$ is the isospin
lowering operator, that turns a neutron into a proton,
has temporal and spatial parts in momentum space:
\begin{eqnarray}
J_{i,{\rm 1b}}^{0}(p^{2}) &=&  1 - g_{A} \, 
\frac{{\bf P} \cdot {\bm \sigma}_{i}}{2 m}  \,, 
\label{Jt} \\
{\bf J}_{i,{\rm 1b}}(p^{2}) &=& g_{A} \, {\bm \sigma}_{i}
+i \kappa_{V} \, \frac{{\bm \sigma}_{i}\times {\bf p}}{2 m}
,
\end{eqnarray}
where ${\bf P}={\bf
p}_i+{\bf p}_i'$, $g_A\approx 1.27$ is the axial constant, and $\kappa_V\approx 4.70$ is the nucleon magnetic moment. We notice that the polar-vector part of the charge operator and the axial-vector part of the current operator are of leading order, while the axial-vector charge and the weak magnetic current are sub-leading, suppressed by the nucleon mass. This leads to an additional suppression of the correction terms by a factor $3-5$. As a result, these corrections are naturally of the order of $0.2-0.4\%$ compared to the leading terms, and can be easily calculated, albeit introduce a dependence on a nuclear model.  

A different source of theoretical corrections are radiative and relativistic corrections to the decay process. In allowed beta decays, decay probability and spectrum are affected by radiative corrections of a few permille \cite{PhysRevC.28.2433,Wilkinson199889,WILKINSON1995497}. One expects similar effects in first-forbidden unique beta decays \cite{Wilkinson199889}. We note that radiative corrections to the spectrum shape are small, as the shape is only sensitive to ratios of matrix elements. This is similar to the $\beta$ asymmetry parameter in allowed $\beta$ decay, where the radiative corrections are of the order of $10^{-4}$ \cite{Yokoo01031976,PhysRevD.46.2090}. Fully relativistic treatment of the spectrum can affect the shape of the spectrum, up to a per-mill level, and will be calculated in the future \cite{JHEP.6.1}.

Thus, these theoretical arguments suggest that the na\"ive expression of Eq.~\ref{Eq:Angle_part_first_forbidden} is expected to be accurate to a few per-milles. Including radiative and recoil corrections is expected to allow a theoretical prediction accurate to $\approx10^{-4}$. 

\begin{figure}[ht]
\begin{center}

\includegraphics[width=0.3\textheight]{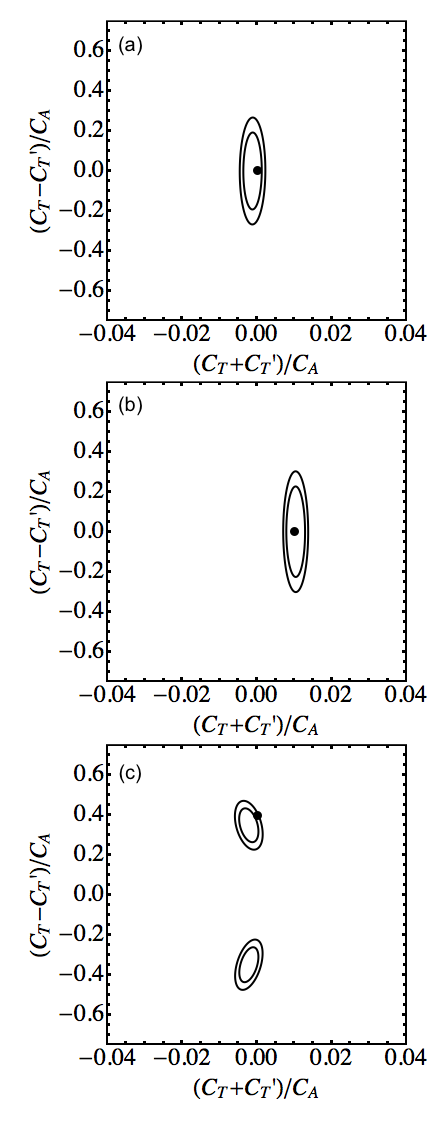}
\end{center}
\caption{\label{Fig:Fits}68\% \& 90\% confidence intervals for the fits to the BSM couplings. See text for details. The bold dot indicates the values of the couplings in the simulation.}
\end{figure}

{\it Experimental sensitivity --}
In order to estimate the experimental sensitivity to BSM physics we simulate a first forbidden unique energy spectrum (containing 10M events), using an endpoint energy of 3.5MeV. 
In order to account for experimental uncertainty in the $\beta$ energy measurement we introduce resolution effects of 20 keV to each of the simulated events. 
The theoretical $\beta$ energy spectrum can be calculated from integration over all angles in Eq. \ref{Eq:Angle_part_first_forbidden}, to obtain:
\begin{eqnarray}
\label{Eq:IntegralSpec}
\frac{dw_{\beta^\mp}}{d\epsilon}\propto\Sigma(\epsilon)\left(2+ 4 \gamma_0 \frac{C_T+C_T'}{C_A}\frac{m_e}{\epsilon}+\frac{\beta}{5}\frac{(a^2-1)\tanh^{-1}(a)+a}{a^2}
\left(1-\frac{|C_T|^2+|C_T'|^2}{|C_A|^2}\right)\right),
\end{eqnarray}
where $a=2k\nu/(k^2+\nu^2)$. A Bayesian analysis, using the JAGS framework~\cite{jags} is then performed on the recorded spectra to extract the endpoint energy, 
overall normalization, and the two BSM parameters. Since $C_T$ and $C_T'$ are fully correlated we parametrize the BSM contribution on the uncorrelated 
parameters $(C_T+C_T')/C_A$ and $(C_T-C_T')/C_A$. Figure~\ref{Fig:Fits} shows the results from 3 such analyses, corresponding to no BSM couplings 
($C_T=C_T'=0$, Fig. \ref{Fig:Fits}(a)),
weak left handed coupling ($C_T/C_A=C_T'/C_A=0.005$, Fig. \ref{Fig:Fits}(b)), and strong right handed coupling ($C_T/C_A=-C_T'/C_A=0.2$, Fig. \ref{Fig:Fits}(c)). 
Note that the case of fully right handed couplings 
cannot be detected in spectrum measurements of allowed decays, since in that case $b=0$. Also note the reduced sensitivity to the right handed case, which arises
due to quadratic dependence on $C_T-C_T'$.

In addition to resolution effects, energy calibration effects may also play a role in the extraction of the spectrum. We performed the same analysis using an energy calibration offset of $\pm$0.5\%, no effect 
was observed for the extracted parameters. We note that energy calibration errors are significant when measuring endpoint energies, where the spectral shape is not parametrized. In the case of a measurement
of the full spectrum, it is the resolution effects (which move events between energy bins in a non-trivial manner) that play an important role. We further note that since the endpoint energy does not depend on the
exact shape of the spectrum an additional constraint may be imposed in the fitting procedure by using measured endpoint energies.

In conclusion, we have proposed the $\beta$-spectrum of unique first-forbidden decay as a novel probe for beyond the standard model couplings in the weak interaction. Analyzing possible systematic uncertainties demonstrates that such studies may surpass the accuracy level of correlation measurements in allowed $\beta$ decays, and, contrary to allowed $\beta$ decays, enable simultaneous extraction of exotic tensor couplings to both right and left handed neutrinos in an uncorrelated 
manner. Of course, the use of a different experiment to study BSM couplings allows the examination of systematic uncertainties in the experiments, particularly essential in precision studies of such minute effects. Our initial study shows that similar results are expected in other forbidden decays.

First-forbidden unique decays are abundant in nature~\cite{Didierjean:1994hb}, vary in $Q$-values, and are amenable for precision spectra measurements, e.g., as studied in antineutrino mass effects on the endpoint of the first forbidden unique decay of $^{187}$Re~\cite{PhysRevC.83.045502,Andreotti2007208}, or in search for hints Lorentz violation in the decay rate \cite{PhysRevLett.111.171601}. Thus, our prediction increases significantly the number of relevant experiments searching BSM effects, and in particular the dimension six tensor type corrections. One such potential measurement which may be carried out is the beta decay of $^{90}$Y, with an endpoint energy of $\sim$2.3 MeV, and may be easily produced via the 
$^{90}$Zr(n,p) reaction in copious amounts.

The work of Y.\ M., M.\ H., and G.\ R. has been supported by the Israeli Science Foundation under ISF grant  139/15 and the Pazy Foundation. A.\ G.\ M. and D.\ G. acknowledge the support of ARCHES.

\end{document}